# Peculiarities of the SEHRS and SERS Spectra of $4,4'$ - Bipyridine Molecule

**Vladimir P. Chelibanov[1], Alexander V. Golovin[2], Aleksey M. Polubotko[3*]**

[1]State University of Information Technologies, Mechanics and Optics, Kronverkskii 49, 197101 Saint Petersburg, RUSSIA  E-mail: Chelibanov@gmail.com

[2]Saint Petersburg State University, Ulianovskaya st. 1, 198504, Saint Petersburg Petrodvoretz RUSSIA E-mail: golovin50@mail.ru

[3] A.F. Ioffe Physico-Technical Institute, Politechnicheskaya 26, 194021 Saint Petersburg, RUSSIA E-mail: alex.marina@mail.ioffe.ru

## Abstract

The SEHRS and SERS spectra of $4,4'$ - Bipyridine are analyzed on the base of the Dipole-Quadrupole theory for two possible geometries of the molecule. It is demonstrated that there appear strong lines caused by vibrations transforming after a unit irreducible representation both for the geometry with $D_2$ and $D_{2h}$ symmetry groups, which may probably describe the symmetry properties of the molecule. Appearance of these lines is associated with a strong quadrupole light – molecule interaction, which arises in nano size regions of sharp roughness of the metal. In addition, there are the lines caused by contributions from both the vibrations transforming after the unit irreducible representations $A$ or $A_g$ and the representations $B_1$ or $B_{1u}$, respectively, which describe transformational properties of the $d_z$ component of the dipole moment, which is perpendicular to the surface for both geometries. This result is associated with a specific geometry of the molecule, when the indicated vibrations can be nearly degenerated and cannot be resolved by the SEHRS and SERS spectra analysis. This issue is in a full compliance with the results of the SEHRS and SERS Dipole-Quadrupole theory.

# Introduction

Surface Enhanced Optical processes are of a great interest, since these processes can be a powerful tool for scientific investigations and are applicable in physics, chemistry and biology. Usually one explains these phenomena by the hypothesis of surface plasmons and a chemical enhancement. However, these approaches do not explain appearance of forbidden lines, which are observed in the spectra of all these processes in molecules with sufficiently high symmetry. This difficulty is overcome in the Dipole-Quadrupole theory, which explains appearance of these lines naturally due to existence of a strong quadrupole light – molecule interaction arising due to strongly inhomogeneous electromagnetic fields, which exist near a rough metal surface[1]. This interaction becomes very strong especially in nano size regions near the sharpest points of the rough metal surface because of a very strong increase of the electric field derivatives $\frac{\partial E_\alpha}{\partial x_\alpha}$. These regions are named as active sites or hot spots because of their strong localization and a strong enhancement of the electric field and its derivatives in them.

In SEHRS the forbidden lines associated with the totally symmetric vibrations, transforming after the unit irreducible representation were observed in pyrazine and phenazine[2-4]. In addition such lines must be observed in other symmetric molecules with sufficiently high symmetry that follows from the results of the Dipole-Quadrupole SEHRS theory[5]. However, there are no experimental data on the SEHRS spectra of other symmetric molecules, which could confirm appearance of this type of forbidden lines. In this work we present theoretical interpretation of some features of the SEHRS and SERS spectra of $4,4^{'}$ - Bipyridine, observed in[6]. The problem is that the symmetry group of the molecule may be under the question and can be $D_2$ or $D_{2h}$. However in[7] it was demonstrated that this molecule most probably belongs to the $D_2$ symmetry group. This result was obtained on the base of the energy minimum fundamental principle for this configuration. Here we shall demonstrate that the most enhanced lines in SEHRS and SERS are caused by vibrations transforming either by the unit irreducible representation, or by the representations $B_1$ and $B_{1u}$, for $D_2$ and $D_{2h}$ symmetry groups respectively. Two last irreducible representations describe transformational properties of the dipole moment component $d_z$ which is perpendicular to the surface. The enhancement of the lines with the unit irreducible representation is associated with the strong quadrupole light–molecule interaction, when the strongest scattering arises due to the quadrupole moments $Q_{xx}, Q_{yy}, Q_{zz}$, or their linear combinations, having a constant sign, which are named as main quadrupole moments. The enhancement of the lines with the $B_1$ and $B_{1u}$ irreducible representations is associated with the scattering via the two main quadrupole moments in SEHRS and one main quadrupole moment in SERS and due to the scattering via the $d_z$ moment, associated with the enhancement of the $E_z$ component of the electric field, which is perpendicular to the surface. However, most of the strongly enhanced lines are associated with both enhancement mechanisms and both types of the scattering since both types of vibrations are nearly degenerate and contribute to the same lines. Here we mean that the lines, caused by these nearly degenerate vibrations cannot be really resolved in the above mentioned experiment[6]. This unique situation strongly differs from the one for pyrazine and phenazine and is associated with a specific geometry of the molecule, which consists of two weekly interacting benzene rings and, therefore, there may exist two nearly degenerated symmetric and antisymmetric vibrational states with very close frequencies.

## Main relations of the Dipole-Quadrupole SEHRS and SERS theories

The Dipole-Quadrupole SEHRS theory was published in[5]. In addition, there is the Dipole-Quadrupole theory of SEIRA and SERS, which is based on the same conceptions. The last one is published in the monograph[1]. Therefore, we present here only some main topics of this theory and reference the reader for more detailed acquaintance to the above mentioned works.

In accordance with the Dipole-Quadrupole theory, there exists a strong quadrupole light – molecule interaction arising in surface fields strongly varying in space near a rough metal surface. This interaction is associated with the quadrupole terms in the light – molecule interaction Hamiltonian for the incident and scattered fields.

$$\widehat{H}_{e-r}^{inc} = |E_{inc}| \frac{(e^* f_e^*)_{inc} e^{i\omega_{inc}t} + (e f_e)_{inc} e^{-i\omega_{inc}t}}{2}, \qquad (1)$$

$$\widehat{H}_{e-r}^{scat} = |E_{scat}| \frac{(e^* f_e^*)_{scat} e^{i\omega_{scat}t} + (e f_e)_{scat} e^{-i\omega_{scat}t}}{2}. \qquad (2)$$

Here $E_{inc}$ and $E_{scat}$ are the incident and scattered electric fields, $\omega_{inc}$ and $\omega_{scat}$ are corresponding frequencies, $\overline{e}$ is the corresponding polarization vector,

$$f_{e\alpha} = d_{e\alpha} + \frac{1}{2E_\alpha} \sum_\beta \frac{\partial E_\alpha}{\partial x_\beta} Q_{e\alpha\beta} \qquad (3)$$

is an $\alpha$ component of generalized vector of interaction of light with the electrons of the molecule and

$$d_{e\alpha} = \sum_i e x_{i\alpha} \text{ and } Q_{e\alpha\beta} = \sum_i e x_{i\alpha} x_{i\beta}, \qquad (4)$$

are an $\alpha$ component of the dipole moment vector and the $\alpha\beta$ component of the quadrupole moments tensor of electrons. Here $x_{i\alpha}$ and $x_{i\beta}$ are the Cartesian coordinates of the $i$ th electron. In accordance with the Dipole – Quadrupole theory, only the terms of the Hamiltonians (1, 2), which are associated with the moments $Q_{xx}, Q_{yy}, Q_{zz}$ and $d_z$ are essential for the strong scattering. These moments are named as main moments $Q_{main}$ and $d_{main}$. Corresponding reasoning one can find in our publications and in[1], for example. The Hamiltonian with the quadrupole and dipole moments can be very significant due to some features of the scattering on these moments, the enhancement of the $E_z$ component of the electric field, which is perpendicular to the surface near the sharp points of the rough surface and very large values of the derivatives $\frac{\partial E_\alpha}{\partial x_\alpha}$ with equal indices, compared with the ones in a free space. In addition the quadrupole interaction is significant because of some properties of matrix elements with the quadrupole moments $Q_{xx}, Q_{yy}$ and $Q_{zz}$.

For such model of roughness as an ideally conductive cone the radial component of the electric field can be expressed as

$$E_r \sim |E_{0,inc}| C_0 \left(\frac{l_1}{r}\right)^\beta . \tag{5}$$

Here $|E_{0,inc}|$ is the amplitude of the incident electric field, $C_0 \sim 1$ is a numerical coefficient, $l_1$ is a characteristic size of the cone, a height for example. $r$ is the radius vector counted from the top of the cone $0 < \beta < 1$ and depends on the cone angle. The enhancement coefficient for the pure dipole enhancement mechanism of SERS and SEHRS is

$$G_d \sim C_0^{2n} \left(\frac{l_1}{r}\right)^{2n\beta} \tag{6}$$

where $n$ is the order of the optical process and $n=2$ for SERS and $n=3$ for SEHRS. The enhancement coefficient for the pure quadrupole enhancement mechanism is

$$G_Q \sim C_0^{2n} \beta^{2n} \left(\frac{B}{2}\right)^{2n} \left(\frac{l_1}{r}\right)^{2n\beta} \left(\frac{a}{r}\right)^{2n} . \tag{7}$$

Here $B \gg 1$ is some numerical coefficient, $a$ is the molecule size. The necessity to introduce the $B$ coefficient arises because the matrix elements of the quadrupole moments with the same indices $Q_{xx}, Q_{yy}$ and $Q_{zz}$ are significantly larger that it was considered in quantum mechanics. This result is associated with the fact that that the above quadrupole moments have a constant sign, while the dipole moments are of a changeable sign. Therefore the matrix elements of these quadrupole moments are larger that it was considered earlier, when one considered that their relation is

$$\frac{\overline{\langle m|Q_{\alpha\alpha}|n\rangle}}{\overline{\langle m|d_\alpha|n\rangle}} \sim Ba \tag{8}$$

and $B \sim 1$. One can see from (6,7) that the dipole interaction with the enhanced component of the electric field $E_z$, which is perpendicular to the surface and the quadruple interaction are strongly enhanced in a small region, when $r \to 0$. For the characteristic sizes of $l_1$ of several nanometers, this region is a nano size one. It should be noted that for a more real roughness, when the top of the cone is not ideally sharp but has a large but a finite curvature, the electric field at the top is not infinity. However in this case the field and its derivatives are strongly enhanced in the nano size region of the top and the dipole and quadrupole interactions are strongly enhanced too. Because of the quadrupole interaction important role in the SEHRS process, various combinations of the dipole and quadrupole moments can contribute to the effect (Figure 1). Then the scattering is determined by several contributions, which arise due to the scattering via various combinations of the dipole and quadrupole moments. Thus the scattering is determined by these contributions, which occur due to these various types of the scattering and the SEHRS cross-section in symmetrical molecules is proportional to the sum of them[5].

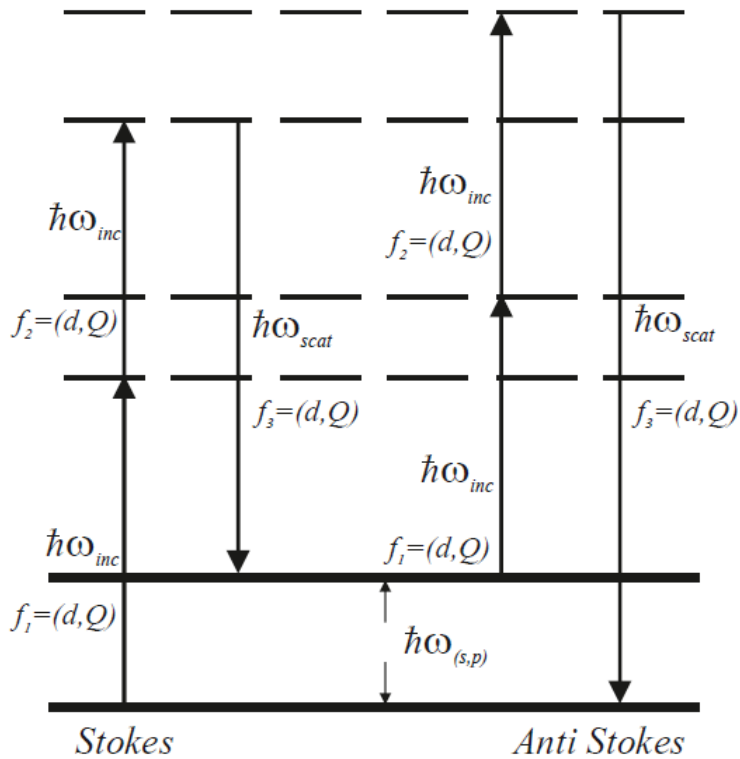

Figure 1. The scattering diagram of SEHRS for the Stokes and Anti Stokes scattering. The virtual absorption and emission can occur via various dipole and quadrupole moments $f_1, f_2$ and $f_3$.

$$\sigma_{SEHRS_S} \sim \sum_p \left| \sum_{f_1, f_2, f_3} T_{(s,p), f_1-f_2-f_3} \right|^2 dO . \tag{9}$$

Here $f_1, f_2$ and $f_3$ stand for various dipole and quadrupole moments, $T_{(s,p), f_1-f_2-f_3}$ means the contribution, which arises due to the scattering via various moments, $dO$ is a solid angle. The indices $(s, p)$ refer to degenerate vibrational modes, where the index $s$ numerates the group of degenerated vibrational states, while $p$ numerates the states inside the group. The contributions $T_{(s,p), f_1-f_2-f_3}$, which we will simply designate further as $(f_1 - f_2 - f_3)$, obey the selection rules

$$\Gamma_{(s,p)} \in \Gamma_{f_1} \Gamma_{f_2} \Gamma_{f_3} \tag{10}$$

where the symbol $\Gamma$ designates the irreducible representation, which describes transformational properties of the $(s, p)$ vibration and the $f_1, f_2$ and $f_3$ moments. There may be several distinct contributions to one line, however the magnitude of these contributions can significantly differ one from another. Therefore, we can sometimes take into account only one, the most enhanced contribution that really defines the value of the scattering cross-section. Since the quadrupole interaction with the main quadrupole moments can be most enhanced, the most enhanced contributions are those, which describe the scattering via three main quadrupole moments: $(Q_{main} - Q_{main} - Q_{main})$. Since the electric field component $E_z$, which is perpendicular to the surface, is strongly enhanced too, the contribution $(Q_{main} - Q_{main} - d_z)$ will be enhanced also, but with a lesser degree. The other enhanced contributions $(Q_{main} - d_z - d_z)$ and $(d_z - d_z - d_z)$ will be enhanced as well but with a lesser degree than the previous two. Since the main quadrupole moments transform after the unit irreducible representation, the most enhanced lines are caused by the vibrations with the unit irreducible representation and with the representation, which describes transformational properties of the $d_z$ moment, which is perpendicular to the surface in accordance with the selection rules (6). The lines, caused by vibrations transforming after the unit irreducible representation, are forbidden in usual Hyper Raman scattering in molecules with sufficiently high symmetry. This situation is implemented in the symmetry groups, where the moment $d_z$ transforms after another irreducible representation, than the unit one. Such lines were observed in pyrazine and phenazine, which belong to the $D_{2h}$ group[2-4] and satisfies this condition. Further we shall need in analysis of the SERS, usual Raman and IR spectra of 4,4'- Bipyridine. Therefore, here we shall briefly describe main results of the theories of these spectra of symmetrical molecules too.

All the features, which concern the influence of the quadrupole light-molecule interaction and the quadrupole and dipole moments are the same in the SEHRS and SERS theories. The SERS cross-section is expressed via the sum of the contributions

$$\sigma_{SERS_S} \sim \sum_p \left| \sum_{f_1, f_2,} T_{(s,p), f_1-f_2} \right|^2 dO . \qquad (11)$$

However every contribution depends only from two dipole and quadrupole moments, in accordance with the order of the process (Figure 2). These contributions obey the selection rules

$$\Gamma_{(s,p)} \in \Gamma_{f_1} \Gamma_{f_2} . \qquad (12)$$

Further we define these contributions simply as $(f_1 - f_2)$. There may be several distinct contributions to one line, such as in SEHRS, however the magnitude of these contributions can significantly differ one from another also. Therefore, we can sometime take into account only one, the most enhanced contribution, which define the value of the cross-section. Since the quadrupole interaction with the main quadrupole moments can be most enhanced, the most enhanced contributions are those, which describe the scattering via two main quadrupole moments: $(Q_{main} - Q_{main})$. Because the electric field component $E_z$, which is perpendicular to the surface is strongly enhanced too, the contribution $(Q_{main} - d_z)$ will be also enhanced. One should note that the lines, caused by the $(Q_{main} - Q_{main})$ are defined by the vibrations with the unit irreducible representations in molecules with a sufficiently high symmetry. However, these lines are not forbidden in SERS, because there is a strongly enhanced contribution $(d_z - d_z)$, which contribute in the lines with the unit irreducible representation. The argument in favour of the existence of the strong quadrupole light-molecule interaction is appearance of strong forbidden lines, caused by the contributions of the $(Q_{main} - d_z)$ type, which are observed in a lot of symmetrical molecules.

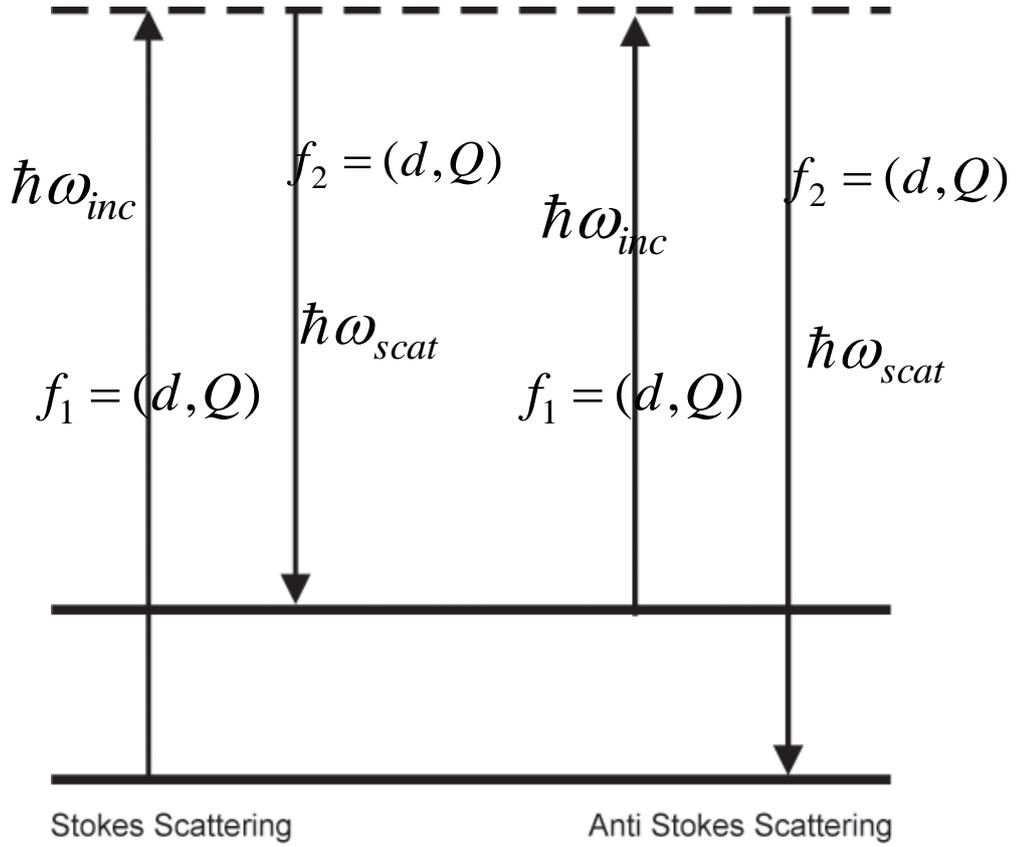

Figure 2. The scattering diagrams of SERS for the Stokes and Anti Stokes scattering. The virtual absorption and emission can occur via various dipole and quadrupole moments $f_1$ and $f_2$.

One should remind that in usual Raman the cross-section and the selection rules formally have the same form as the expressions (7) and (8). However, the $f_1$ and $f_2$ moments are only of the dipole type. Therefore, in a usual Raman scattering there are the lines, caused by vibrations with the unit irreducible representation such as in SERS, which are not forbidden. In usual IR there are only the lines, caused by the vibrations transforming as the dipole moments.

# Analysis of the SEHRS spectrum of 4,4'-Bipyridine for the geometry with the $D_2$ symmetry group

Let us next consider the regularities of the SEHRS and SERS spectra of 4,4'-Bipyridine. The SEHRS spectrum of this molecule was published in[6]. The problem is that the geometry of this molecule is under a question, as was mentioned above. The authors of [7] determined the geometry which consists from two connected benzene rings, with the connection between carbon atoms and where the opposite carbon and hydrogen atoms are substituted by nitrogen atoms. They came to conclusion that the benzene rings do not lie in the same plane and are turned with respect to each other on the angle of 38.7 degrees (Figure 3).

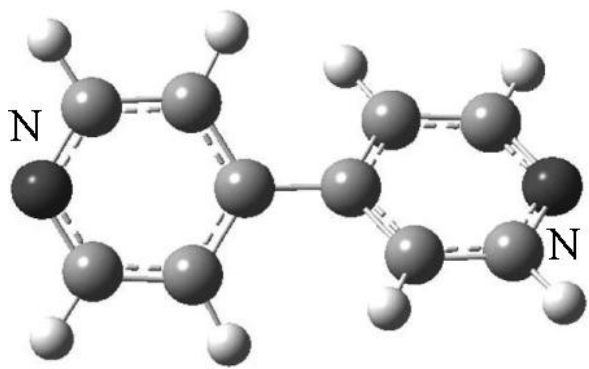

Figure 3. The structure of the 4, 4'-Bipyridine molecule with the $D_2$ symmetry group. The benzene rings are turned with respect to each other on the angle of 38.7 degree.

The symmetry group of the molecule in this case is $D_2$. Similar opinion is expressed in[8]. We made a full geometry optimization and determined the frequencies and symmetry of vibrations of the 4, 4'-Bipyridine molecule using the hybrid functional B3LYP with the 6-31G(d) basic set in the Gaussian 03 program[9]. We would like to stress that our calculations with the program Gaussian 03 result in the same conclusion. However, in literature researchers often consider that the molecule has a plain geometry and belongs to the $D_{2h}$ symmetry group. The difference between the energy of the molecule with the $D_{2h}$ symmetry group and with the $D_2$ group, calculated with the above program is ~0.108 eV that is sufficiently large, compared with $kT \sim 0.026$ eV at $T = 300K$. ($k$ is the Boltzmann constant). Therefore, in this paragraph we assume that the molecule belongs to the $D_2$ symmetry group, which has four irreducible representations. For correct determination of the symmetry of vibrations we need to know orientation of the molecule with respect to the coordinate system. It was chosen is such a manner that the $z$ axis passes via two nitrogen atoms. Since 4, 4'- Bipyridine adsorbs mainly vertically and is connected with the substrate via the nitrogen atom, then the $z$ axis coincides with the enhanced $E_z$ component of the electric field, which is perpendicular to the surface. We calculated vibrational wavenumbers and determined the symmetry of vibrations of the molecule by the program Gaussian 03. The SEHRS and SERS spectra of 4,4'-Bipyridine is presented on Figure 4 and the calculated wavenumbers and the assignment are collected in Table 1. In the first column the experimental wavenumbers of the SEHRS vibrational spectrum, taken from[6] and the qualitative estimation of the intensity of spectral lines are presented. The second the third and the forth columns present experimental values of the wavenumbers of the SERS, usual Raman and IR lines, taken from [5], [8] and Internet respectively. Other columns present calculated vibrational wavenumbers, which are most close to the experimental values and are the most probable assignment of these vibrations. One should note that experimental values, obtained from various literature sources can differ one from another because of different experimental conditions in various experiments. However this difference is not crucial for our main conclusions.

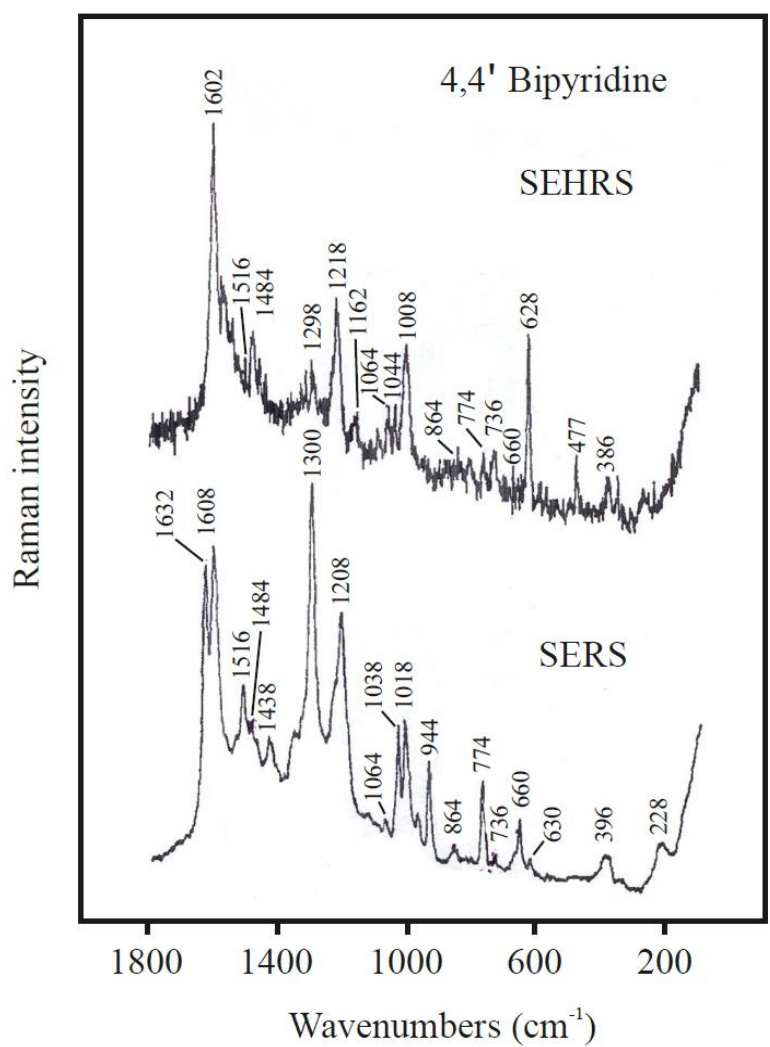

Figure 4. The SEHRS and SERS spectra of 4,4'-Bipyridine[3]

Table 1. Assignment of the SEHRS, SERS, usual Raman and IR lines of 4, 4'-Bipyridine for $D_2$ symmetry group

| Wavenumber $cm^{-1}$ experiment | | | | The most close calculated values of wavenumbers in $cm^{-1}$ and the most possible assignment. | | Additional possible assignment in the $D_2$ symmetry group | |
|---|---|---|---|---|---|---|---|
| SEHRS | SERS | Raman | IR | $A$ | $B_1$ | $B_2$ | $B_3$ |
| 386 w | 396 w | 388 w | No data | $A$ (383) | $B_1$ (377) | | |
| 477 m | | | | | | | ? |
| | | | 499 s | | | | $B_3$ (500) |
| | | 574 w | 570 s | | | $B_2$ (570) | |
| 628 s | 630 vw | | | | | | ? |
| 660 vw | 660 m | 661 m | | | | | $B_3$ (658) |
| | | | 674 w | | | $B_2$ (670) | |
| 736 w | 736 vw | | 742 m | $A$ (743) | | $B_2$ (743) | |
| 774 w | 774 m | 767 m | | | | | ? |
| | | | 807 vs | | | | $B_3$ (797) |
| 864 vw | 864 vw | 875 w | | | $B_1$ (865) | | |
| | | | 881 w | | | | ? |
| | 944 m | | | | | $B_2$ (954) | $B_3$ (952) |
| 1008 m | 1018 m | 1011 vs | 990 s | $A$ (985) | $B_1$ (979) | | |
| 1044 w | 1038 m | | 1040 m | | $B_1$ (1029) | | |
| 1064 w | 1064 vw | 1080 w | 1076 m | $A$ (1073) | $B_1$ (1068) | | |
| 1162 w | | | | | | | ? |
| 1218 m | 1208 m | 1224 m | 1219 s | $A$ (1222) | $B_1$ (1218) | | |
| 1298 m | 1300 vs | 1296 vs | | $A$ (1282) | | | |
| | 1438 w | | | | | | $B_3$ (1419) |
| 1484 w | 1487 vw | | 1489 s | | $B_1$ (1487) | | |
| 1516 vw | 1516 m | 1514 m | | $A$ (1507) | | | |
| | | | 1532 m | | | $B_2$ (1543) | |
| 1602 vs | 1608 s | 1606 s | 1602 vs | $A$ (1605) | $B_1$ (1600) | | |
| | 1632 s | 1624 s | | | | | ? |

One can see that the most probable vibrations, which determine the lines of the SEHRS spectrum are those with the unit irreducible representation $A$ and the irreducible representation $B_1$, which determine transformational properties of the $d_z$ moment. The corresponding experimental values of wavenumbers for these vibrational modes are 386, 1008, 1064, 1218 and 1602 $cm^{-1}$. Besides there are two lines with the unit irreducible representation $A$ only at $1298 cm^{-1}$ and $1516 cm^{-1}$. Existence of these lines and sufficiently strong intensity of the line with the wavenumber $1298 cm^{-1}$ points out the existence of the strong quadrupole light – molecule interaction. Therefore we cannot ignore and must consider that this interaction contributes significantly in the first mentioned set of the lines too. Besides, a good proof that these lines are caused by the strong quadrupole interaction only is the absence of IR lines with these wavenumbers and their presence in SERS and in usual Raman that confirms their belonging to the unit irreducible representation $A$. The lines caused by the totally symmetric vibrations with the unit irreducible representation are forbidden in usual HRS in molecules with sufficient high symmetry. However, in molecules with $D_2$ symmetry group they are formally allowed. This situation occurs since there is a contribution in these lines caused by the scattering via $d_x, d_y$ and $d_z$ moments (the $(d_x - d_y - d_z)$ contribution). However it cannot be strong, but is really very weak and must be close to zero since it is determined by tangential components of the electric field $E_x$ and $E_y$, which are parallel to the surface and are nearly equal to zero because of a nearly ideal conductivity of the metal surface. This result is valid for the contributions from the molecules, which are adsorbed in the first layer. However it is well known that the molecules, adsorbed in the second and other layers enhance the SEHRS signal significantly lower (The first layer effect[12]). Therefore, the above contribution is determined mainly by the scattering from the first layer and must be very weak. However, the contribution $(Q_{main} - Q_{main} - Q_{main})$ is huge and determines appearance of the strong lines with the unit irreducible representation $A$. This situation is very similar to the one for phenazine and pyrazine[2-4] and corroborate our theoretical results about appearance of forbidden lines in the SEHRS spectra[5,10].

The lines with very small difference of the wavenumbers having the symmetry $A$ and $B_1$ follow in pairs for a most part of the vibrational modes. It is necessary to note that there are a lot of such pairs with close wavenumbers in our calculations of the vibrational spectrum. This effect is apparently associated with the specific geometry of the molecule. Two benzene rings are connected with the $C-C$ bond, which changes the vibrational frequencies of the benzene rings only slightly. Therefore, there are two vibrational states, symmetric and antisymmetric, which must have close vibrational frequencies and various symmetry $A$ and $B_1$. Similar situation was described in[10-11] where the existence of such pairs with very close calculated wavenumbers and various symmetry was noted for the molecule *trans*-1,2-bis(4-peridyle)ethylene. This molecule with a $C_{2h}$ symmetry group also consists of two slightly modified benzene rings, which are weakly connected via a chain of carbon atoms. The interaction between the rings can be weak that results in appearance of two nearly degenerate symmetric and antisymmetric states, which are linear combinations of the states of the rings. For 4,4'-Bipyridine, because of the very close values of the wavenumbers with the difference, which cannot be resolved sometimes in the experiments, both vibrations contribute in the same line in fact and therefore may not be distinguished from experiment unambiguously. In accordance with the Dipole-Quadrupole theory, appearance of the strong lines with the $B_1$ symmetry is associated with the $(Q_{main} - Q_{main} - d_z)$ types of the scattering. The fact that the lines with the wavenumbers 386, 1008, 1064, 1218 and 1602 $cm^{-1}$ can have $(Q_{main} - Q_{main} - d_z)$ contribution is confirmed by the result that there are IR lines with very close wavenumbers for all these lines. Thus analysis of the SEHRS spectrum of 4,4'- Bipyridine for the indicated geometry demonstrates that there are the lines, caused both by the vibrations with the irreducible representations $A$ and $B_1$ with very close wavenumbers associated with strong quadrupole and dipole light-molecule interactions and both types of the interactions manifest in experimental SEHRS, SERS, usual Raman and IR spectra. In addition there are several lines in the spectra, which may be caused by the vibrations with the $B_1$ symmetry only. They are ones with the wavenumbers 864, 1044 and 1487 $cm^{-1}$. This fact does not contradict to our theory since these lines exist in the SEHRS, SERS and IR spectra simultaneously. Moreover, the usual Raman signal is absent, or very slight (for the line 864 $cm^{-1}$) for these lines that corresponds to the theory of this process. We must notice that we do not know detailed experimental conditions, when these spectra are obtained. In addition, there may be large uncertainty in assignment due to its approximate character. Therefore, we do not explain some details of these spectra with other wavenumbers. This remark concerns the SEHRS lines 477, 628, 774, 1162 $cm^{-1}$ and the SERS lines with close wavenumbers, the line 1632 $cm^{-1}$ in SERS and the line 881 $cm^{-1}$ in the IR spectrum. All these cases are designated by the symbol "?" in the last column of Table 1. It is necessary to note that the lines with $B_2$ and $B_3$ irreducible representations are caused by the contributions in the SEHRS spectra, which contain tangential components of the electric field $E_x$ and $E_y$, which are nearly equal to zero at the surface. Therefore these SEHRS lines are weak or absent at all that one can see from Table 1. The same can be said concerning the SERS lines with this symmetry. Sufficiently strong intensity of the SERS lines at 660 and 944 $cm^{-1}$ may be associated with adsorption of the molecules on the atomic scale roughness, like steps. The orientation of the molecule with respect to the surface can be arbitrary and the $E_z$ perpendicular component can have all enhanced components in the coordinate system, associated with the molecule. Therefore this fact may be the reason of these sufficiently strong intensities.

However in any case one can state that in the case of the $D_2$ symmetry group there are the lines in the SEHRS and SERS spectra, which can be caused simultaneously by the vibrations with $A$ and $B_1$ irreducible representations. Existence of these lines well agrees with the Dipole-Quadrupole theory.

## Analysis of the SEHRS and SERS spectra of 4,4'-Bipyridine for the geometry with the $D_{2h}$ symmetry group

Since there is an opinion, that 4, 4'-Bipyridine belongs to the $D_{2h}$ symmetry group, we performed calculations of the vibrational wavenumbers and determinate the symmetry of vibrations for this case too (Table 2). There are the lines with the wavenumbers $1298 cm^{-1}$ and $1516 cm^{-1}$, which are caused by the vibrations with the unit irreducible representation $A_g$ which are forbidden in usual HRS (Hyper Raman Scattering). This fact also indicates the existence of the strong quadrupole light-molecule interaction in this case such as in the case when we considered that the molecule belongs to the $D_2$ symmetry group. In addition there are strong lines, caused by the vibrations with the unit irreducible representation $A_g$ and the representation $B_{1u}$, which describes transformational properties of the $d_z$ moment. These lines at 1008, 1064, 1218 and 1602 $cm^{-1}$ have strictly similar calculated wavenumbers as the lines with the irreducible representation $A$ and $B_1$ in the case of the $D_2$ symmetry, except of the line 386 $cm^{-1}$, which has no such assignment.

Table 2. Assignment of the SEHRS, SERS, usual Raman and IR lines of 4, 4'-Bipyridine for a plane geometry and $D_{2h}$ symmetry group

| Wavenumber $cm^{-1}$ experiment | | | | The most close calculated values of wavenumbers in $cm^{-1}$ and the most possible assignment. | | Possible assignment in the $D_{2h}$ symmetry group for other irreducible representations | |
|---|---|---|---|---|---|---|---|
| SEHRS | SERS | Raman | IR | $A_g$ | $B_{1u}$ | | |
| 386 w | 396 w | 388 w | No data | | | $B_{1g}$ (378) | |
| 477 m | | | | | | | ? |
| | | | 499 s | | | | ? |
| | | 574 w | 570 s | | | $B_{2g}$ (568) | IR ? |
| 628 s | 630 vw | | | | | | ? |
| 660 vw | 660 m | 661 m | | | | $B_{3g}$ (657) | |
| | | | 674 w | | | $B_{2u}$ (674) | |
| 736 w | 736 vw | | 742 m | $A_g$ (742) | | $B_{3u}$ (727) | |
| 774 w | 774 m | 767 m | | | | | ? |
| | | | 807 vs | | | | $B_{3u}$ (792) |
| 864 vw | 864 vw | 875 w | | | | $B_{1g}$ (857) | |
| | | | 881 w | | | | ? |
| | 944 m | | | | | $B_{2g}$ (941) | $B_{3u}$ (949) |
| 1008 m | 1018 m | 1011 vs | 990 s | $A_g$ (990) | $B_{1u}$ (1026) | | |
| 1044 w | 1038 m | | 1040 m | | $B_{1u}$ (1063) | | |

| 1064 w | 1064 vw | 1080 w | 1076 m | $A_g$ (1089) | $B_{1u}$ (1063) | | |
|---|---|---|---|---|---|---|---|
| 1162 w | | | | | | | ? |
| 1218 m | 1208 m | 1224 m | 1219 s | $A_g$ (1241) | $B_{1u}$ (1223) | | |
| 1298 m | 1300 vs | 1296 vs | | $A_g$ (1269) | | | |
| | 1438 w | | | | | $B_{3g}$ (1421) | |
| 1484 w | 1487 vw | | 1489 s | | $B_{1u}$ (1489) | | |
| 1516 vw | 1516 m | 1514 m | | $A_g$ (1517) | | | |
| | | | 1532 m | | | $B_{2u}$ (1545) | |
| 1602 vs | 1608 s | 1606 s | 1602 vs | $A_g$ (1603) | $B_{1u}$ (1602) | | |
| | 1632 s | 1624 s | | | | | ? |

In addition, there are strong lines with the $B_{1u}$ irreducible representation only, which describe transformational properties of the $d_z$ moment. They are the lines with $1484$ and $1044 cm^{-1}$. Thus the situation is absolutely similar to the situation with the $D_2$ symmetry group. It is necessary to note that our assignment to the irreducible representations $A_g$ and $B_{1u}$ well coincides with the assignment, which was used for interpretation of the IR and usual Raman spectra of 4,4' Bipyridine, published in[13] that strongly confirms our ideas. However, because of approximate character of the calculations there are some disagreements in interpretation of the origin of some lines in the SEHRS, SERS, usual Raman and IR spectra. The lines with the assignment difficulties are designated by the sign "?" in the last column of Table 2. Here it is designated also what lines of the spectra is difficult to assign. There are 8 such cases. This number is more that the number of such cases in the $D_2$ group of the molecule, which is equal to 6. The lines with other symmetry than $A_g$ and $B_{1u}$ correspond to the symmetry of the lines in the $D_2$ group with the following compliance ($B_{1u}, B_{1g} \rightarrow B_1, B_{2u}, B_{2g} \rightarrow B_2$, $B_{3u}, B_{3g} \rightarrow B_3$). However, there is some exclusion for the SEHRS line 736 $cm^{-1}$. In general, the assignment in the $D_{2h}$ symmetry group is worse than in the $D_2$ group from our point of view. However, the main result that the SEHRS and SERS lines (1008, 1064, 1218 and 1602 $cm^{-1}$) can be assigned both to the unit irreducible representation $A_g$ and the irreducible representation $B_{1u}$ remains the same. Thus, sometimes there is no unambiguous assignment for the vibrational modes because of the close values of the calculated wavenumbers of the vibrations with $A_g$ and $B_{1u}$ symmetry, which follow in pairs. However, existence of the lines with these irreducible representations demonstrates that even for the plane geometry, which apparently is less probable, the SEHRS and the SERS spectra of 4, 4' Bipyridine can be described by the Dipole-Quadrupole theory.

## Conclusion

Thus, our analysis of the SEHRS and SERS, usual Raman and IR spectra of 4, 4-Bipyridine demonstrates existence of sufficiently strong lines, caused by totally symmetric vibrations, transforming after the unit irreducible representation in both possible geometries of the molecule, belonging to the $D_2$ and $D_{2h}$ symmetry groups. Because of very small value of the contribution of the $(d_x - d_y - d_z)$ type, which contribute in the lines with the $A$ symmetry (in the molecule with the $D_2$ symmetry group), we can say that appearance of these lines is a manifestation of the strong quadrupole light-molecule interaction in this system. There are several lines caused by vibrations with the $B_1$ or $B_{1u}$ irreducible representations, which indicate existence of the strong dipole light-molecule interaction associated with the enhancement of the $E_z$ component of the electric field, which is perpendicular to the surface. In addition, there are several lines, which can be caused both by the nearly degenerated vibrations with $A$ and $B_1$ symmetry in the $D_2$ group, or with $A_g$ and $B_{1u}$ symmetry in the $D_{2h}$ group, which apparently cannot be resolved in the experiment. The intensity of these lines can be formed both by the strong $(Q_{main} - Q_{main} - Q_{main})$ and $(Q_{main} - Q_{main} - d_z)$ types of the scattering in SEHRS, or by $(Q_{main} - Q_{main})$ and $(Q_{main} - d_z)$ scattering types in SERS. The peculiarity of the SEHRS spectra with respect to the one for the pyrazine molecule[3,4], is the fact that several lines must be assigned to both irreducible representations $A$ and $B_1$ or $A_g$ and $B_{1u}$. This situation occurs due to specific geometry of the molecule, which consists of two benzene rings, which slightly affect the vibrations of each other and form symmetric and anti symmetric vibrational states with very close frequencies. In conclusion, we should say that in our opinion the geometry with the $D_2$ symmetry group is more probable. This conclusion arises because there is the minimum of the total energy of the molecule for this configuration and the assignment is more correct for this case.